# Partition Function Zeros in Non-compact QED

A. Ali Khan and I. Barbour

Department of Physics and Astronomy, University of Glasgow, Glasgow G12 8QQ, U.K.

The lowest zeros of the lattice partition function for non-compact QED are found in the complex fermion mass plane on $6^4$, $8^4$ and $10^4$ lattices at intermediate values of the coupling. The scaling of the low lying zeros with lattice size is analysed.

## 1. Introduction

We present preliminary results for the behaviour of the lowest zeros in the complex fermion mass plane of the lattice partition function of 4-dimensional non-compact QED with four flavours of staggered fermions. Results on $6^4$, $8^4$ and $10^4$ lattices are presented for values of the inverse coupling $\beta$ in a region where there is a possible tricritical point in the (zero temperature) infinite volume limit.

The work presented here is the first part of an ongoing study of the scaling behaviour of the zeros as a function of increasing lattice volume.

## 2. Method

An ensemble of configurations is generated using a Hybrid Monte Carlo algorithm, choosing the molecular dynamics parameters to keep the length of each molecular dynamics trajectory $\sim 0.8$ and the acceptance rate between 0.65 and 0.8.

Measurements are made on configurations separated by approximately 3 or 4 units of molecular dynamics time.

The lattice partition function is defined as

$$Z(m) = \frac{\int dA \det(M + m\mathbb{1}) e^{-S_g}}{\int dA \det(M + m_0 \mathbb{1}) e^{-S_g}}, \quad (1)$$

where $m_0$ is the update mass, $M$ is the fermion matrix at zero fermion mass, $m$, and $S_g$ is the standard non-compact gauge field action. The eigenvalues of $M$ are found using the Lanczos algorithm [1] to tridiagonalise $M$. The tridiagonal form is then diagonalized using Sturm sequences. Since $iM$ is hermitian, this method gives the eigenvalues without reorthogonalization and its consequent storage demands. Accurate determination of all the eigenvalues enables accurate determination of the coefficients of the characteristic polynomial (each weighted with respect to the determinant at the update mass) [2].

These coefficients, averaged over the ensemble, give the polynomial expansion in $m^2$ for the partition function

$$Z = \sum_{n=0}^{L^4/2} e^{a_n} m^{2n} \quad (2)$$

on an $L^4$ lattice.

We choose the exponential parametrization of the coefficients because they vary within a large range of values (on a $6^4$ lattice between about 0 and $e^{550}$, on a $10^4$ lattice between about 0 and $e^{4400}$) and grow linearly with the volume.

The zeros of this polynomial are the Lee-Yang zeros in the complex mass plane.

Since the eigenvalues of $M$ appear as complex conjugate imaginary pairs, $Z$ is a polynomial in $m^2$ and its coefficients are necessarily positive. The fact that the coefficients vary grossly in magnitude gives rise to technical difficulties:
a) in their determination
b) in finding the roots of the polynomial.
These difficulties have been overcome and the methods adopted will be described in a subsequent publication.

## 3. Results

For non-compact QED with staggered fermions on a $6^4$ lattice for $0.16 \leq \beta \leq 0.22$, the zeros near the physical mass region ($m \geq 0$ and real) are imaginary. This suggests that any phase tran-sition will only occur at zero fermion mass, as expected. Note that this signal differs from that in compact QED [3], where the lowest zeros at a coupling of $\beta = 0.885$ are found to be complex with non-zero real part.

Errors on the zeros are not shown since they are difficult to estimate in a rigorous manner. The statistical errors can be estimated via a bootstrap technique. This gives an error of about n% on the n'th zero (with n=1 corresponding to the lowest zero). A related systematic error will also arise from limited ensemble sampling. We estimate this error to be comparable with the statistical error.

It is also relevant to note that the $a_n$ are approximating to a continuous function of $n$. The equation

$$\sum_{n=0}^{L^4/(2k)} e^{a_{kn}} m^{2kn} = 0 \qquad (3)$$

($k$ integer), becomes, if the $a_n$ become continuous in the infinite volume limit, for $V$ large

$$\frac{1}{k} \int_0^{V/2} e^{a(x)} m^{2x} dx = 0 \qquad (4)$$

and its roots are the same as those of $Z$. That is, using equation 3 with $k = 1, 2, 3, \ldots$, we find the same roots near the physical region. This signal for the continuity of $a_n$ grows stronger with increasing lattice volume.

Figure 1 shows the lowest zeros plotted in sequence. It is easily seen that in the infinite volume limit and in the limit $m \to 0$, the condensate is given by $\rho(0)$, the density of zeros at the origin. For large enough lattices, this density corresponds to the normalized gradient of the lines at fixed $\beta$ at the lowest zeros. Hence from the measurements on a $6^4$ lattice we estimate the transition to occur between $\beta_c \sim 0.185$ and $0.19$.

On larger lattices, we observe a change in the distribution of the zeros.

For $\beta = 0.195$ on the $8^4$ lattice and for $\beta \leq 0.200$ on the $10^4$ lattice, the lowest zeros are imaginary (at $\beta = 0.195$ on the $10^4$ lattice some of the lowest zeros fluctuate very slightly away from the imaginary axis but this could be due to inadequate ensemble sampling). At $\beta = 0.210$ on the $10^4$ lattice, only the lowest of the small zeros is imaginary and at $\beta = 0.220$ all the zeros have gained a small real part. Note that only the zeros close to the origin are determined with relatively small error.

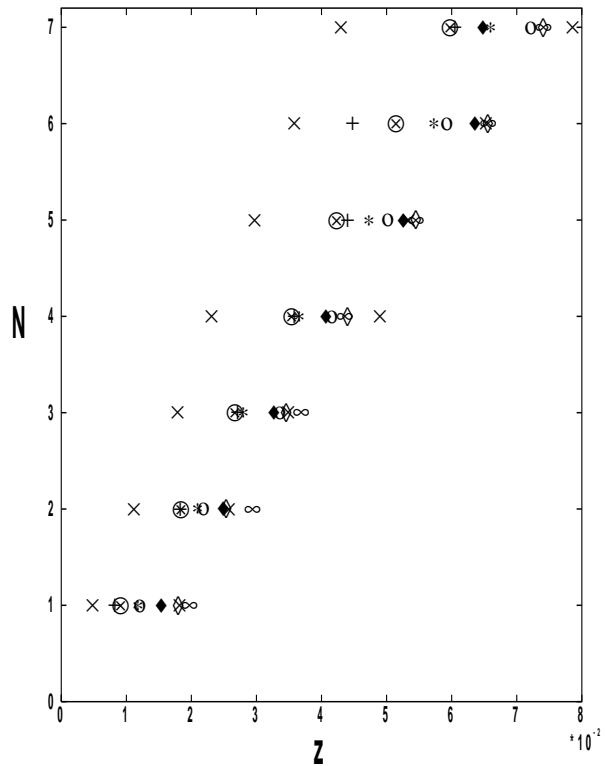

Figure 1. *The lowest zeros on the $6^4$ lattice plotted in sequence. The crosses correspond to $\beta = 0.160$, crossed circles to $\beta = 0.185$, pluses to $\beta = 0.190$, asterisks to $\beta = 0.195$, open circles to $\beta = 0.200$, filled diamonds to $\beta = 0.205$, double circles to $\beta = 0.210$, open diamonds to $\beta = 0.215$ and the crosses on the right hand side to $\beta = 0.220$.*

Because we did not observe this change in the distribution within the above range of $\beta$ on the $6^4$ lattice, we performed a simulation at $\beta = 0.25$ on a $6^4$ lattice. After a couple of hundred measurements the lowest zero obtained a small real part, but it became complex again after $\sim 1000$ measurements.

Figure 2 shows the lowest zeros on the $10^4$ lattice plotted in sequence for $\beta < 0.21$. At $\beta = 0.185$ the transition clearly looks first order. At $\beta = 0.195$ it appears to be close to first order. At that $\beta$ we investigated the scaling behaviours of the lowest zero and of the separations between the first and second, and the second and third zero. The separations scale with respect to the $8^4$ and $10^4$ lattices as $V^\alpha$ with $\alpha$ just greater than $-1$, again consistent with a behaviour which is slightly weaker than first order. The lowest zero on the other hand scales with $\alpha < -1$. At $\beta = 0.2$, the data from the $10^4$ lattice indicate that $\rho(0) = 0$.

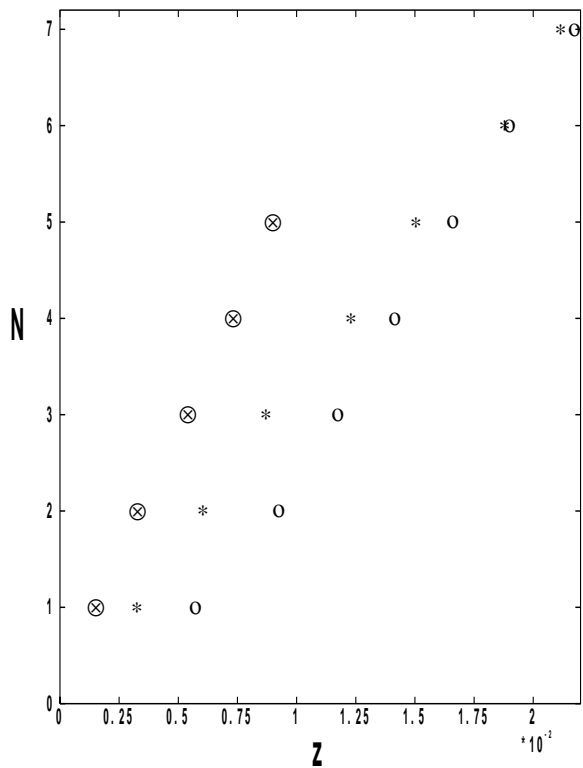

Figure 2. *The lowest zeros on the $10^4$ lattice plotted in sequence for $\beta < 0.21$. The crossed circles correspond to $\beta = 0.185$, the asterisks to $\beta = 0.195$ and the open circles to $\beta = 0.200$.*

However, in view of the large finite size effects, these conclusions may change when larger volumes are simulated.

It is clearly important to study this change in the distribution of the zeros on larger lattices.

The migration of the zeros from the imaginary axis could be a signal for a change in the critical behaviour of the system. On the other hand it could be a result of finite size effects, i.e. our lattice is too small for the onset of finite size scaling. If not, it could be that via finite size scaling the real parts of the lowest zeros scale to the imaginary axis if the critical mass is zero.

## 4. Conclusions

The fact that there is a tricritical point [4] appears to be reflected in the rapid increase in magnitude of the lowest zeros in that region and in the change in their distribution. We have evidence that finite size effects are large in the weak coupling sector.

We intend to perform simulations on larger lattices to investigate these points in greater detail.

### Acknowledgements

We would like to thank John Kogut, Ely Klepfish, Gaetano Salina and Gerrit Schierholz for very useful comments and suggestions. IB would like to thank the Illinois and Jülich (HLRZ) Groups for allowing him to use their HMC programs. Partial support by EC contract CHRX − CT92 − 0051 is acknowledged.